\documentclass[prl,twocolumn,lengthcheck,superscriptaddress]{revtex4-1}

\usepackage{amsmath,braket}
\usepackage{amsthm}
\usepackage{amssymb}
\usepackage{graphicx}
\usepackage{hyperref}
\usepackage{color}
\usepackage[textsize=tiny]{todonotes}
\usepackage{tikz}
\usetikzlibrary{calc}

\usetikzlibrary{positioning,shapes.geometric,decorations.markings,arrows,knots,hobby} 
\usepackage[many]{tcolorbox}
\usepackage{pgfplots}
\usepgfplotslibrary{patchplots}
\usepgfplotslibrary{fillbetween}

\usepackage{dsfont}

\newcommand{\tr}{\operatorname{tr}}

\theoremstyle{definition}

\theoremstyle{remark}

\def\eqbox#1{\tcboxmath[colback=light,arc=0cm,frame hidden,enhanced]{#1}}

\definecolor{accent1}{HTML}{A65B6F}
\definecolor{light}{HTML}{EBEDF2}
\definecolor{medium}{HTML}{D2D4D9}
\definecolor{dark}{HTML}{7297A6}
\definecolor{accent2}{HTML}{A6554E}

\begin{document}

\title{No Free Lunch for Quantum Machine Learning}
\author{Kyle Poland}

\email{kyle.poland0310@gmail.com}
\author{Kerstin Beer}

\email{kerstin.beer@itp.uni-hannover.de}
\author{Tobias J.\ Osborne}
\affiliation{Institut f\"ur Theoretische Physik, Leibniz Universit\"at Hannover, Appelstr. 2, 30167 Hannover, Germany}

\begin{abstract}
	The ultimate limits for the quantum machine learning of quantum data are investigated by obtaining a generalisation of the celebrated \emph{No Free Lunch} (NFL) theorem. We find a lower bound on the quantum risk (the probability that a trained hypothesis is incorrect when presented with a random input) of a quantum learning algorithm trained via pairs of input and output states when averaged over training pairs and unitaries. The bound is illustrated using a recently introduced QNN architecture.
\end{abstract}

\maketitle

\emph{Machine learning} (ML), particularly as applied to \emph{deep neural networks} via the \emph{backpropagation algorithm}, has brought about enormous technological and societal change \cite{Goodfellow2016,Nielsen2015,Jordan2015,murphyMachineLearningProbabilistic2012}. Myriad applications now range the full gamut from image analysis and self-driving cars, through to the placement and removal of customized advertisements \citep{selfdrivingcars1, selfdrivingcars2, image_analysis, advertisement_placement, advertisement_removal}. Purely classical ML continues to enjoy rapid progress, however, the advent of quantum computation promises a bevy of powerful new tools and generalisations.

We are now witnessing the experimental arrival of large-scale quantum information processors \cite{Google2019}. Such \emph{noisy intermediate-scale quantum devices} (NISQ) \cite{preskillQuantumComputingNISQ2018a} have ushered in the \emph{quantum information era} and present critical new challenges and opportunities for theoretical physics. A most pressing challenge is how to cope with the imminent ubiquity of \emph{quantum data} when quantum devices commence the routine production of complicated entangled states involving $50$ or more qubits. The characterisation of such states goes far beyond practical tomography; instead, a natural tool to process the coming surge in quantum data will be quantum devices themselves via \emph{quantum machine learning} (QML).

The nascent field of QML \cite{Biamonte2017,cilibertocarloQuantumMachineLearning2018,schuldQuantumMachineLearning2017} carries
great promise for the discovery of quantum learning algorithms by exploiting quantum
analogues of the artificial neural network (ANN) architecture \cite{Goodfellow2016,murphyMachineLearningProbabilistic2012,Nielsen2015} to carry out the learning of quantum
data. Classically, ANNs are superbly well-adapted for classification problems via supervised and unsupervised learning of training data and there is optimism that quantum analogues will enjoy comparable success. Several quantum architectures have been considered so far, including, variational quantum circuits \cite{farhiQuantumAlgorithmsFixed2017} and a variety of neural network-like architectures \cite{wanQuantumGeneralisationFeedforward2017,farhiClassificationQuantumNeural2018,killoranContinuousvariableQuantumNeural2019}. Recently, a promising candidate artificial quantum neural network architecture (QNN) was introduced \cite{beerTrainingDeepQuantum2020}. Initial investigations have shown that these QNNs are well adapted to both supervised \cite{beerTrainingDeepQuantum2020} and unsupervised learning tasks \cite{bondarenkoQuantumAutoencodersDenoise2019}. 

Understanding the ultimate limits for quantum learning devices and methods is a key priority, a goal central to \emph{quantum learning theory} (QLT) \cite{arunachalamGuestColumnSurvey2017,gammelmarkQuantumLearningMeasurement2009,sasakiQuantumTemplateMatching2001,sasakiQuantumLearningUniversal2002,sentisQuantumLearningQuantum2012,monrasInductiveSupervisedQuantum2017a}. The field of QLT has enjoyed steady progress during the past years, amassing a variety of key results particularly characterising the limits for quantum devices to learn classical data, encoded in special quantum states, and also for classical devices to learn quantum states. There has been comparatively less progress on the problem of characterising the ultimate limits for the learning of ``fully'' quantum data by quantum devices themselves.

The goal of this paper is to progress quantum learning theory for general quantum data by generalising a celebrated result in classical learning theory, the \emph{No Free Lunch} (NFL) theorem \cite{wolpertNo1997} to the quantum setting. More precisely, we demonstrate an optimal lower bound on the probability that a quantum information processing device --- modelled as a unitary process trained with quantum examples --- incorrectly acts on a randomly chosen input. This bound provides the ultimate limit for quantum machine learning and thus furnishes us with a practical metric to determine the functioning of QML architectures and algorithms. We illustrate the obtained bound using the recently introduced QNN architecture of \cite{beerTrainingDeepQuantum2020}. This result is related to work on the optimal quantum learning of unitary operations, as introduced in \cite{bisioOptimalQuantumLearning2010a}, which considered the storage and later retrieval of unknown quantum processes.

{\color{accent1} \paragraph{Preliminaries}\hspace{-1em}}.---The classical NFL theorem \cite{wolpertNo1997} establishes that an optimization algorithm exhibiting elevated
performance for one class of problems must perform worse for another class. There are many formulations of the NFL theorem; we prefer a version adapted to learning algorithms (we follow, e.g., the presentation given in  \cite{Wolf2018}). 

To explain the classical NFL theorem we introduce some notation. Let $X$ and $Y$ be two finite sets, called the \emph{input} and \emph{output} sets, respectively. The goal is to determine a \emph{hypothesis} $h:X\rightarrow Y$ for an unknown function $f:X\rightarrow Y$, given access only to a \emph{training subset} $S\subset X\times Y$ consisting of a list of \emph{training examples}: $S = \{(x_j,y_j)\,|\, x_j\in X, y_j\in Y, j = 1,2, \ldots n\}$, where $y_j = f(x_j)$ is the correct output given input $x_j$. That is, $h$ should obey $h(x_j) = y_j = f(x_j)$ for all $j=1,2,\ldots, n$. For such a learning task to be nontrivial we assume that $n < |X|$ (for otherwise the hypothesis $h$ would already be completely determined and there would be nothing to predict). To quantify how well a given hypothesis performs at modelling $f$ we introduce the \emph{risk} $R_f(h)$ as follows:
\begin{equation}
	R_f(h) \equiv \mathbb{P}[h(x)\not=f(x)],
\end{equation}
i.e., as the probability, with respect to the uniform distribution of $x\in X$ over $X$, that $h$ gives an incorrect answer. We can now quote the NFL theorem as the lower bound
\begin{equation} \label{eq:original_NFL_statement}
	\eqbox{
	\mathbb{E}_f\left[\mathbb{E}_S\left[R_f(h_S)\right]\right]\ge \left(1-\frac{1}{|Y|}\right)\left(1-\frac{n}{|X|}\right),}
\end{equation}
where $\mathbb{E}_f$ denotes the average with respect to the uniform distribution over all possible functions from $X$ to $Y$, $\mathbb{E}_S$ denotes the uniform average over all possible training sets with $n$ elements, and $h_S$ denotes an information-theoretically optimal hypothesis given the training data $S$. Intuitively the NFL theorem tells us that if a learning algorithm performs better at predicting $f$ for some problem instance then there are other problem instances where the algorithm will perform worse.

{\color{accent1} \paragraph{The quantum NFL theorem}\hspace{-1em}}.---The natural quantum analogue of the NFL theorem applies to quantum devices which are optimised to reproduce quantum training examples, presented as pairs of inputs to, and outputs from, a quantum device. To describe this setting we first replace the input and output sets $X$ and $Y$ above with \emph{input} and \emph{output Hilbert spaces} $\mathcal{H}_{\text{in}}$ and $\mathcal{H}_{\text{out}}$, respectively. We write $d=\dim(\mathcal{H}_{\text{in}})$ and $d'=\dim(\mathcal{H}_{\text{out}})$ for their corresponding dimensions (these dimensions play the role of $|X|$ and $|Y|$ in the classical setting). In the quantum case the object playing the role of the unknown function $f$ is an unknown \emph{unitary process} $U$. This process models an uncharacterised quantum device. A \emph{training set} is then a list $S\subset \mathcal{H}_{\text{in}}\otimes \mathcal{H}_{\text{out}}$:
\begin{equation}
	\{|\phi_j\rangle|\psi_j\rangle\,|\, |\phi_j\rangle\in\mathcal{H}_{\text{in}}, |\psi_j\rangle\in\mathcal{H}_{\text{out}}, j=1,2,\ldots, n\}
\end{equation}
of pairs of input and output states. The training pairs are assumed \emph{ideal} and \emph{realisable}, meaning that all of the pairs obey $|\psi_j\rangle = U|\phi_j\rangle$ for all $j=1,2,\ldots, n$. The goal is then to determine a hypothesis unitary $V$ which reproduces the action of the unknown unitary $U$ on inputs from $S$; since we are interested in the ultimate limits on quantum learning we demand that $V$ reproduces $U$ exactly:
\begin{equation}
	V|\phi_j\rangle = |\psi_j\rangle = U|\phi_j\rangle, \quad j=1,2,\ldots,n.
\end{equation}
To assess how well the hypothesis performs in reproducing the action of $U$ we introduce the \emph{quantum risk} as the trace-norm distance between the outputs of $U$ and $V$ applied to the same input, averaged over all pure states (see, e.g., \cite{monrasInductiveSupervisedQuantum2017a} for a discussion of the risk in the quantum setting)
\begin{equation}\label{eq:quantumrisk}
	\begin{split}
		R_{U}(V) &\equiv \int d|\psi\rangle \,\| U|\psi\rangle\langle \psi|U^\dag - V|\psi\rangle\langle \psi|V^\dag \|_1^2, \\
		&\equiv 1-\int d|\psi\rangle \, |\langle\psi|U^\dag V|\psi\rangle|^2, \\
		&\equiv 1-\frac{1}{d(d+1)}\left(d+|\tr(U^\dag V)|^2\right),
	\end{split}
\end{equation}
where $\|A\|_1 \equiv \tfrac12\tr|A|$ is the trace norm \cite{nielsenQuantumComputationQuantum2000} and the integral is over pure states induced by a Haar-measure-distributed unitary $W$ applied to a fiducial product state $d|\psi\rangle \equiv dW|0\rangle$ \cite{duistermaatLieGroups2000,haydenAspectsGenericEntanglement2006a}. This quantity, just as in the classical case, represents the probability that, when confronted with a random input $|\psi\rangle$, the hypothesis fails to (i.e., may be detected to) reproduce the action of $U$. (See the supplementary material for an elementary derivation of this equation.)

The setting we have in mind is as follows: imagine that an untrusted complex quantum device acting as a unitary $U$ on a large number of qubits is purchased from a purveyor. The goal is to characterise (or, more nefariously, reverse-engineer) the device. We imagine that the quantum device may be reproducibly applied to input states of our choosing. Given such a device we may easily prepare training pairs $\{|\phi_j\rangle\otimes (U|\phi_j\rangle)\}_{j=1}^N$ of input-output pairs. Exploiting the training pairs we may train an architecture $V$ given by, e.g., a QNN, to learn the action of the quantum device. To do so we make measurements of, or coherently interact with, the training pairs, yielding an approximation to the action of $U$. The ultimate limit for how well we can train $V$ is then given by assuming that $V$ perfectly reproduces the action of $U$ on $S$. 

We can now quote the quantum analogue of the NFL theorem, which applies to the quantum risk, uniformly averaged over all problem instances $U$, with respect to Haar measure, and all sets $S$ of $n$ training pairs:
\begin{equation}\label{eq:quantumNFL}
	\eqbox{\mathbb{E}_U[\mathbb{E}_{S}[R_U(V)]] \ge 1-\frac{1}{d(d+1)}(n^2+d+1).}
\end{equation}
We compare below the classical and quantum NFL theorems. In doing this, one should keep in mind that a classical function $f$ may readily be many to one, and hence not invertible, so that having determined $f$ on some subset of the inputs one gains no additional information about the action of $f$ on the complement of this subset. However, a unitary process $U$ is always invertible, so that once we have determined $U$ on some subspace, we already have the additional information that $U$ takes the complementary subspace to a complementary subspace of the output. In this way one might argue that one should properly compare the quantum NFL theorem with a classical NFL-type theorem for \emph{invertible} functions. We have derived such a bound, which reads $\mathbb{E}_f\left[\mathbb{E}_S\left[R_f(h_S)\right]\right]\ge  1 - \frac{n+1}{|X|}$, in the supplementary material. This bound behaves very similarly to the standard classical NFL theorem, apart from a slightly different slope, which reflects the additional information supplied by the assumption the function $f$ is invertible.

The quantum NFL provides an apparently stronger lower bound than its classical counterpart. Intuition for this might be extrapolated from the case of a single qubit: given a single training example $(|\phi\rangle, U|\phi\rangle)$ one might expect that we have completely determined the action of $U$ because the complement of the subspace $\mathcal{K}$ spanned by $|\phi\rangle$ must be mapped to the complementary subspace determined by $U|\phi\rangle$. However, this is not the case as there is still the freedom to choose a phase that $U$ applies to $|\psi\rangle \in \mathcal{K}$. When evaluating the risk averaged over Hilbert space this freedom affects the action of $U$ on almost all inputs because the average is taken over all superposition inputs $c_0|\phi\rangle + c_1|\psi\rangle$. By contrast, for a classical function on a binary alphabet $\{0,1\}$, once we have a single training pair we already know the action $f$ on half of the inputs and are reduced to the problem of guessing a binary output, for which we will be correct $50\%$ of the time. The probability of being incorrect is then only $25\%$.

{\color{accent1} \paragraph{Proof of the quantum NFL theorem}\hspace{-1em}}.---The argument for the bound Eq.~(\ref{eq:quantumNFL}) proceeds as follows. We must average the quantum risk expression
\begin{equation*}
	R_{U}(V) = 1-\frac{1}{d(d+1)}\left(d+ |\tr(U^\dag V)|^2 \right)
\end{equation*}
over all training sets $S$ and all possible unitaries $U$. (Note that $V$ \emph{implicitly depends} on $U$ in a potentially very complicated way: it is the best guess for $U$ given the information afforded by the training set $S$.) The first average is trivial as the quantum risk only depends on the number of elements of the training set $S$. The second average requires that we evaluate the following integral
\begin{equation}\label{eq:qrisk2}
	\int dU\, R_{U}(V) = \frac{d}{d+1} - \frac{1}{d(d+1)} \int dU |\tr(U^\dag V)|^2.
\end{equation}
The integral on the RHS may be evaluated according to the following strategy. 
The hypothesis $V$ is a unitary which acts identically to $U$ on the training set $S$. However, by linearity we automatically learn that $U$ and $V$ agree on the subspace $\mathcal{H}_{S} \equiv \text{span}(S)$. While we have no information about the action of $U$ on the subspace $\mathcal{H}^{\perp}_{S}$ complementary to $\mathcal{H}_S$, we do still know that $U$ is unitary, which delivers additional information via the defining quadratic constraints of a unitary operator.  

To understand the interplay between the unitarity constraints and the information supplied by the training set we consider the unitary $U^\dag V$. Thanks to the training set we have the following block decomposition with respect to the direct sum decomposition $\mathcal{H}_{\text{in}} = \mathcal{H}_S\oplus \mathcal{H}_{S}^{\perp}$:
\begin{equation}
	U^{\dagger} V =
		\left( \begin{array}{c|c}
			\mathds{1}_n 	& A\\
			\hline 	
			B			& W
\end{array} \right),
\end{equation}
where $\mathds{1}_n$ is the $n$-dimensional identity on the subspace $\mathcal{H}_S$, and $A$, $B$, and $W$ are $n\times (d-n)$, $(d-n)\times n$, and $(d-n)\times (d-n)$ block matrices, respectively. The unitarity constraints on $U^\dag V$ now force $A=B=\mathbf{0}$ (this is because the norm of each row and column of a unitary must be equal to $1$), so that we obtain the following block decomposition
\begin{align}
U^{\dagger} V =
\left( \begin{array}{c|c}
\mathds{1}_n 	& \mathbf{0}\\
\hline
\mathbf{0}			& W
\end{array} \right) = \mathds{1}_n \oplus W,
\end{align}
where, further, $W$ is now a $(d-n)$-dimensional \emph{unitary}.

As the trace of the $n$-dimensional identity equals $n$ we can decompose the trace of $U^\dag V$ into a sum of traces over $\mathcal{H}_S$ and $\mathcal{H}^{\perp}_{S}$, respectively: we thus obtain
\begin{equation}
\begin{split}
|\tr(U^{\dagger} V)|^2 &= |\tr_{\mathcal{H}_S}(U^{\dagger} V) + \tr_{\mathcal{H}_S^{\perp}}(U^{\dagger} V)|^2  \\
&= |n + \tr_{\mathcal{H}_S^{\perp}}(U^{\dagger} V)|^2  \\
&= n^2 + 2 n \Re(\tr_{\mathcal{H}_S^{\perp}}(U^{\dagger} V)) + |\tr_{\mathcal{H}_S^{\perp}}(U^{\dagger} V)|^2  \\
&= n^2 + 2 n \Re(\tr(W)) + |\tr(W)|^2. 
\end{split}
\end{equation}

The only information we have left about the block matrix $W$, having exploited the unitary constraints, is that it is unitary. Since the action of $W$ is completely undetermined the only strategy left open to us is to guess $W$ randomly with respect to Haar measure on the unitary group $\mathcal{U}(d-n)$. Thus, the average over $U$ is reduced to performing an average of $W$ over the unitary group $\mathcal{U}(d-n)$:
\begin{multline}
	\int dU |\tr(U^\dag V)|^2 =  \\ \int dW \left(n^2 + 2 n \Re(\tr(W)) + |\tr(W)|^2\right)
\end{multline}
Because the second integrand on the RHS is linear in $W$ it vanishes. The third integrand has the value $1$ (see the supplementary material for an elementary derivation). Thus we obtain
\begin{equation}
	\int dU |\tr(U^\dag V)|^2 =  n^2 +1.
\end{equation}
Substituting this into the RHS of Eq.~(\ref{eq:qrisk2}) yields the desired lower bound.

{\color{accent1} \paragraph{Case study: quantum NFL for QML via quantum neural networks}\hspace{-1em}}.---In this section we illustrate the quantum NFL theorem in the case of a recently introduced quantum neural network architecture \cite{beerTrainingDeepQuantum2020}. We studied QNNs which have two input and two output neurons, corresponding to maps $\mathcal{E}:\mathcal{B}(\mathbb{C}^2\otimes \mathbb{C}^2)\rightarrow \mathcal{B}(\mathbb{C}^2\otimes \mathbb{C}^2)$ from states of two qubits to two qubits. We refer to \cite{beerTrainingDeepQuantum2020} for extensive details of the QNN architecture and numerical methods for their optimisation. For the investigation here we may simply regard QNNs as a variational class of maps which may be optimised, e.g., via gradient descent, to optimise the 
output fidelity, averaged over the training data:
\begin{equation}
	C = \frac{1}{n} \sum_{j=1}^n \langle \phi_j|U^\dag \mathcal{E}(|\phi_j\rangle\langle\phi_j|) U|\phi_j\rangle.
\end{equation}

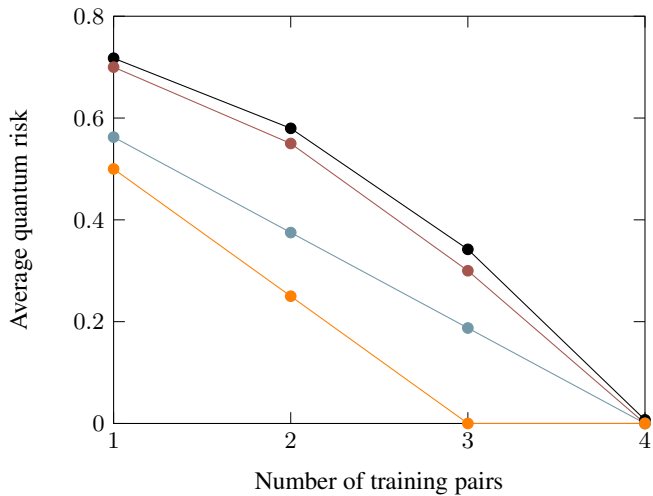
\begin{figure}
	\begin{tikzpicture}[scale=1]
	\begin{axis}[xlabel={Number of training pairs},ylabel={Average quantum risk},width=\columnwidth, height=7cm, tick style={color=black}, domain=0:4,xtick={1,2,3,4}, xmin = 1, xmax = 4, ymin = 0, ymax = 0.8]
	\addplot[color=black,mark=*] table[x=n, y=numeric,  col sep=comma] {2_2_network_raw.txt};
	\addplot[color=accent2,mark=*] table[x=n, y=nflQ,  col sep=comma] {2_2_network_nflQ.txt};
	\addplot[color=dark,mark=*] table[x=n, y=nflC,  col sep=comma] {2_2_network_nflC.txt};
	\addplot[color=orange,mark=*] table[x=n, y=nflCinv,  col sep=comma] {2_2_network_nflCinv.txt};
	\end{axis}
	\end{tikzpicture}
	\caption{Average quantum risk for a QNN when learning an unknown two-qubit unitary (black). Also shown is the lower bound supplied by the quantum NFL theorem ({\color{accent2} brown}), the classical NFL theorem ({\color{dark} blue}), and the classical NFL theorem for invertible functions ({\color{orange} orange}).}
\label{fig:risk}
\end{figure}

To compare with the quantum NFL bound we first chose a unitary $U$ uniformly at random from Haar measure, then $n=1,2,3,4$ training pairs uniformly at random, and then we optimised the cost function $C$. Finally, we evaluated the quantum risk by randomly choosing input states. Forming the empirical average yielded an estimate for the average quantum risk. The results are plotted in Fig.\ref{fig:risk}. 
As one may observe, agreement is good, with the QNN ansatz yielding results close to achieving the quantum NFL bound. The remaining discrepancy is likely due to the fact that the QNNs were not trained to $100\%$ average fidelity. Note that due to the requirement that we perform \emph{three} empirical averages to evaluate the average quantum risk the numerical overhead for obtaining these results is substantial, ruling out verification for larger systems. 

{\color{accent1} \paragraph{Conclusions}\hspace{-1em}}.---We have contributed to quantum learning theory for general quantum data by obtaining a generalisation of the celebrated no free lunch theorem. We did this by obtaining a 
lower bound on the averaged quantum risk, the probability that a quantum information processing device -- modelled as a unitary process trained with quantum examples -- incorrectly acts on a randomly chosen input. This bound was obtained exploiting identities for integrals over the unitary group with respect to the Haar measure and provides the ultimate limit
for quantum machine learning. One may regard the quantum NFL bound as a metric to determine the functioning of QML architectures and algorithm; we illustrated the bound using a QNN architecture, obtaining good agreement with the lower bound.

{\color{accent1} \paragraph{Acknowledgments}\hspace{-1em}}.---Helpful correspondence
and discussions with Dmytro Bondarenko, Lorenzo Cardarelli, Polina Feldmann, Alexander Hahn, Amit Jamadagni, Maria Kalabakov, Sebastian Kinnewig, Roger Melko, Laura Niermann, Simone Pfau, Deniz E. Stiegemann, and E. Miles
Stoudenmire are gratefully acknowledged. Thanks also to Marvin Schwiering, whose Python Code was
used for numerical verification. This work was supported, in part, by the
DFG through SFB 1227 (DQ-mat), the RTG 1991, and funded by the Deutsche Forschungsgemeinschaft (DFG, German Research Foundation) under Germany’s Excellence Strategy – EXC-2123 QuantumFrontiers – 390837967.

\widetext
\appendix

\section{Supplementary material}

\subsection{Haar measure integral identities for the unitary group}

In this appendix we collect some useful identities for integrals over the unitary group. The objective is to present arguments exploiting only elementary linear algebra, calculus, and the defining properties of the Haar measure.  This approach is based on conversations with Aram Harrow and Matt Hastings.

We write an integral over the unitary group $\mathcal{U}(d)$ of $d\times d$ matrices of a (matrix-valued) function $f(U)$ on $\mathcal{U}(d)$ with respect to Haar measure as
\begin{equation}
	I = \int dU\, f(U).
\end{equation}
The defining property of the Haar measure is left- (respectively, right-) invariance with respect to shifts via multiplication: let $V\in \mathcal{U}(d)$ be a fixed unitary, then:
\begin{equation}
	\int dU\, f(UV) = \int d(U'V^\dag)\, f(U') = \int dU'\, f(U').
\end{equation}

The first identity we prove here is 
\begin{equation}
	\eqbox{S_2 \equiv \int dU \, U^\dag \otimes U = \frac1d \textsc{swap}.}
\end{equation}
To achieve this we note that for any hermitian operator $X$ the operator $S_2$ obeys
\begin{equation}
	S_2 = \int dU \, (U^\dag e^{-i\epsilon X}) \otimes (e^{i\epsilon X}U) = (\mathds{1}\otimes e^{i\epsilon X})S_2(e^{-i\epsilon X}\otimes \mathds{1}),
\end{equation}
where $\epsilon>0$ is taken to be infinitesimally small. Expanding to first order in $\epsilon$ and cancelling gives us
\begin{equation}
	0 = i\epsilon(\mathds{1}\otimes X)S_2 - i\epsilon S_2(X\otimes \mathds{1}),
\end{equation}
i.e.,
\begin{equation}
	S_2(X\otimes \mathds{1}) = (\mathds{1}\otimes X)S_2.
\end{equation}
Since this is true for any hermitian operator we choose $X$ to be each of a Hilbert-Schmidt orthonormal hermitian operator basis $\lambda^{\alpha}$, $\alpha = 0,1, \ldots, d^2-1$, with $\tr(\lambda^{\alpha}\lambda^{\beta}) = \delta^{\alpha}$. Choosing $X=\lambda^\alpha$, multiplying on the right by $\lambda^\alpha\otimes \mathds{1}$ and summing over $\alpha$ gives
\begin{equation}\label{eq:s2identity}
	\sum_{\alpha}S_2(\lambda^\alpha\lambda^\alpha\otimes \mathds{1}) = \sum_{\alpha}(\mathds{1}\otimes \lambda^\alpha)S_2(\lambda^\alpha\otimes \mathds{1}).
\end{equation}

We now note that 
\begin{equation}
	\textsc{swap} = \sum_{\alpha} \lambda^\alpha\otimes \lambda^\alpha.
\end{equation}
In terms of a tensor-network diagram this identity reads\\
\begin{center}
	\begin{tikzpicture}[scale=0.8] 
	\draw (0,0) to[out=0,in=240](1,.5)to[out=70,in=180](2,1);
	\draw[line width=3pt, white] (0,1)to[out=0,in=90](1,.5)to[out=270,in=180](2,0);
	\draw (0,1)to[out=0,in=110](1,.5)to[out=290,in=180](2,0);
	\node[] at (2.5, .35)   (a) {$\displaystyle =\sum_\alpha$};
	\node[draw,circle,scale=.9] at (4, 0)   (b) {$\lambda^\alpha$};
	\node[draw,circle,scale=.9] at (4, 1)   (c) {$\lambda^\alpha$};
	\draw (3,0)--(b) -- (5,0);
	\draw (3,1)--(c) -- (5,1);
	\end{tikzpicture}
\end{center}
Connecting the outputs, we find the identity
\begin{center}
	\begin{tikzpicture}[scale=0.8] 
	\node[] at (0, -.16)   (a) {$\displaystyle\sum_\alpha$};
	\node[draw,circle,scale=.9] at (1.5, 0)   (b) {$\lambda^\alpha$};
	\node[draw,circle,scale=.9] at (3, 0)   (c) {$\lambda^\alpha$};
	\draw (.5,0)--(b) -- (c) --(4,0);
	\begin{scope}[xshift=5.05cm,yshift=0cm,scale=.75]
	\node at (-.35, -.16)   (d) {$\displaystyle=\sum_\alpha$};
	\draw (1,1)--(3,1);
	\draw[dark] (3,0) arc(-90:90:.5);
	\draw[dark] (1,0)-- (3,0);
	\draw[dark] (1,0) arc(90:270:.5);
	\draw (1,-1)--(3,-1);
	\node[draw,circle,scale=.9,fill=white] at (2, 1)   (e) {$\lambda^\alpha$};
	\node[draw,circle,scale=.9,fill=white] at (2, -1)   (f) {$\lambda^\alpha$};
	\end{scope}
	\begin{scope}[xshift=8cm]
	\node at (0, 0)   (d) {$=$};
	\end{scope}
	\begin{scope}[xshift=8.5cm,yshift=-.35cm,scale=1.25]
	\draw[dark] (0,0) arc(90:270:.25) to[out=0,in=240](1,0)to[out=70,in=180](2,.5) arc(-90:90:.25);
	\draw (0,0) to[out=0,in=240](1,.5)to[out=70,in=180](2,1);
	\draw[line width=3pt, white] (0,1)to[out=0,in=110](1,.5)to[out=290,in=180](2,0);
	\draw (0,1)to[out=0,in=110](1,.5)to[out=290,in=180](2,0);
	\end{scope}
	\begin{scope}[xshift=12cm]
	\node at (0, 0)   (d) {$=d  \,\mathds{1}$};
	\end{scope}
	\end{tikzpicture}
\end{center}
Representing Eq.~(\ref{eq:s2identity}) as 
\begin{center}
	\begin{tikzpicture}[scale=0.8] 
	\node[] at (0, 0)   (a) {$\displaystyle \sum_\alpha$};
	\begin{scope}[xshift=.5cm,yshift=-.5cm]
	\draw (0,1)--(4,1);
	\draw (0,0)--(4,0);
	\node[draw,circle,scale=.9,fill=white] at (2, 1)   (b) {$\lambda^\alpha$};
	\node[draw,circle,scale=.9,fill=white] at (3, 1)   (c) {$\lambda^\alpha$};
	\node[draw,ellipse,scale=.9,fill=white,minimum height=1.5cm] at (1, .5)   (c) {$S_2$};
	\end{scope}
	\begin{scope}[xshift=5.25cm]
	\node at (0, 0)   (d) {$\displaystyle =\sum_\alpha$};
	\end{scope}
	\begin{scope}[xshift=6cm,yshift=-.5cm]
	\draw (0,1)--(4,1);
	\draw (0,0)--(4,0);
	\node[draw,circle,scale=.9,fill=white] at (1, 0)   (b) {$\lambda^\alpha$};
	\node[draw,ellipse,scale=.9,fill=white,minimum height=1.5cm] at (2, .5)   (c) {$S_2$};
	\node[draw,circle,scale=.9,fill=white] at (3, 1)   (c) {$\lambda^\alpha$};
	\end{scope}
	\end{tikzpicture}
\end{center}
and substituting the above identities we find
\begin{center}
	\begin{tikzpicture}[scale=0.8] 
	\node[] at (0, 0)   (a) {$d$};
	\begin{scope}[xshift=.5cm,yshift=-.5cm]
	\draw (0,1)--(2,1);
	\draw (0,0)--(2,0);
	\node[draw,ellipse,scale=.9,fill=white,minimum height=1.5cm] at (1, .5)   (c) {$S_2$};
	\end{scope}
	\begin{scope}[xshift=3cm]
	\node at (0, 0)   (d) {$=$};
	\end{scope}
	\begin{scope}[xshift=4cm,yshift=-.75cm,scale=.5]
	\draw (-1,3)--(3,3);
	\draw (1,0)--(6,0);
	\node[draw,ellipse,scale=.9,fill=white,minimum height=1.5cm] at (2, 1.5)   (c) {$S_2$};
	\draw[line width=3pt, white] (1,2)--(4,2);
	\draw[dark] (1,2)--(4,2);
	\draw[dark] (4,2) to[out=0,in=240](5,2.5)to[out=70,in=180](6,3);
	\draw[line width=3pt, white] (1,1)--(3,1);
	\draw[dark] (1,1)--(3,1);
	\draw[line width=3pt, white] (3,1) arc(-90:90:1);
	\draw[dark] (3,1) arc(-90:90:1);
	\draw[dark] (1,1) arc(90:270:.5);
	\draw[dark] (-1,0) to[out=0,in=240](0,1)to[out=70,in=180](1,2);
	\end{scope}
	\end{tikzpicture}
\end{center}
Exploiting the integral representation of $S_2$ as
\begin{center}
	\begin{tikzpicture}[scale=0.8] 
	\begin{scope}[xshift=.5cm,yshift=-.5cm]
	\draw (0,1)--(2,1);
	\draw (0,0)--(2,0);
	\node[draw,ellipse,scale=.9,fill=white,minimum height=1.5cm] at (1, .5)   (c) {$S_2$};
	\end{scope}
	\begin{scope}[xshift=3.25cm]
	\node at (0, 0)   (d) {$\displaystyle=\int du$};
	\end{scope}
	\begin{scope}[xshift=4cm,yshift=-.5cm]
	\draw (0,1)--(2,1);
	\draw (0,0)--(2,0);
	\node[draw,circle,scale=.8,fill=white,minimum height=.9cm] at (1, 1)   (b) {$U^\dagger$};
	\node[draw,circle,scale=.8,fill=white,minimum height=.9cm] at (1, 0)   (b) {$U$};
	\end{scope}
	\end{tikzpicture}
\end{center}
and wiring together the outputs:
\begin{center}
	\begin{tikzpicture}[scale=0.8] 
	\begin{scope}[scale=.75]
	\draw (1,1)--(3,1);
	\draw (1,-1)--(3,-1);
	\node[draw,ellipse,scale=.9,fill=white,minimum height=1.5cm,text depth=.7cm] at (2, 0)   (c) {$S_2$};
	\draw[line width=3pt, white] (3,0) arc(-90:90:.5);
	\draw[line width=3pt, white] (1,0)-- (3,0);
	\draw[line width=3pt, white] (1,0) arc(90:270:.5);
	\draw[dark] (3,0) arc(-90:90:.5);
	\draw[dark] (1,0)-- (3,0);
	\draw[dark] (1,0) arc(90:270:.5);
	\end{scope}
	\begin{scope}[xshift=4.25cm,scale=.75]
	\node[] at (-.75, 0)   (a) {$\displaystyle=\int du$};
	\draw (1,1)--(3,1);
	\draw[dark] (3,0) arc(-90:90:.5);
	\draw[dark] (1,0)-- (3,0);
	\draw[dark] (1,0) arc(90:270:.5);
	\draw (1,-1)--(3,-1);
	\node[draw,circle,scale=.8,fill=white,minimum height=.9cm] at (2, 1)   (e) {$U^\dagger$};
	\node[draw,circle,scale=.8,fill=white,minimum height=.9cm] at (2, -1)   (f) {$U$};
	\end{scope}
	\begin{scope}[xshift=7.75cm]
	\node at (-.5, 0)   (a) {$=$};
	\draw (0,0)--(2,0);
	\end{scope}
	\end{tikzpicture}
\end{center}

Putting this together we obtain
\begin{center}
	\begin{tikzpicture}[scale=0.8] 
	\begin{scope}[xshift=-2cm]
	\draw (0,1)--(2,1);
	\draw (0,0)--(2,0);
	\node[draw,ellipse,scale=.9,fill=white,minimum height=1.5cm] at (1, .5)   (c) {$S_2$};
	\end{scope}
	\begin{scope}[xshift=.8cm,yshift=.5cm]
	\node[] at (0,0)   (a) {$\displaystyle=\frac{1}{d}$};
	\end{scope}
	\begin{scope}[xshift=1.8cm]
	\draw (0,0) to[out=0,in=240](1,.5)to[out=70,in=180](2,1);
	\draw[line width=3pt, white] (0,1)to[out=0,in=90](1,.5)to[out=270,in=180](2,0);
	\draw (0,1)to[out=0,in=110](1,.5)to[out=290,in=180](2,0);
	\end{scope}
	\end{tikzpicture}
\end{center}
Thus we conclude that
\begin{equation}
	S_2 = \frac{1}{d}\textsc{swap}.
\end{equation}
This result allows us to evaluate integrals such as
\begin{equation}
	\int dU \, |\tr(U)|^2 = \tr(S_2) = \frac{1}{d}\tr(\textsc{swap}) = 1.
\end{equation}

Our next discussion concerns the operator
\begin{equation}
	S_4 = \int dU \, U^\dag\otimes U^\dag\otimes U\otimes U 
\end{equation}
which we represent graphically via 
\begin{center}
	\begin{tikzpicture}[scale=0.8] 
	\begin{scope}[xshift=.5cm,yshift=-.5cm]
	\draw (0,3)--(2,3);
	\draw (0,2)--(2,2);
	\draw (0,1)--(2,1);
	\draw (0,0)--(2,0);
	\node[draw,ellipse,scale=.9,fill=white,minimum height=3cm] at (1, 1.5)   (c) {$S_4$};
	\end{scope}
	\begin{scope}[xshift=3.25cm,yshift=1cm]
	\node at (0, 0)   (d) {$\displaystyle=\int du$};
	\end{scope}
	\begin{scope}[xshift=4cm,yshift=-.5cm]
	\draw (0,3)--(2,3);
	\draw (0,2)--(2,2);
	\draw (0,1)--(2,1);
	\draw (0,0)--(2,0);
	\node[draw,circle,minimum height=.9cm,scale=.8,fill=white] at (1, 3)   (b) {$U^\dagger$};
	\node[draw,circle,minimum height=.9cm,scale=.8,fill=white] at (1, 2)   (b) {$U^\dagger$};
	\node[draw,circle,minimum height=.9cm,scale=.8,fill=white] at (1, 1)   (b) {$U$};
	\node[draw,circle,minimum height=.9cm,scale=.8,fill=white] at (1, 0)   (b) {$U$};
	\end{scope}
	\end{tikzpicture}
\end{center}

As for $S_2$, if we make an infinitesimal change of variables $U\mapsto e^{i\epsilon X} U$, we obtain the following equation, to first order in $\epsilon$:
\begin{equation}
	S_4 (X\otimes\mathds{1}\otimes\mathds{1}\otimes\mathds{1}) + S_4 (\mathds{1}\otimes X\otimes\mathds{1}\otimes\mathds{1}) = (\mathds{1}\otimes\mathds{1}\otimes X\otimes\mathds{1}) S_4 + (\mathds{1}\otimes\mathds{1}\otimes\mathds{1}\otimes X) S_4.
\end{equation}
Choosing $X = \lambda_\alpha$, multiplying on the right by $\lambda_\alpha \otimes\mathds{1}\otimes\mathds{1}\otimes\mathds{1}$, and summing over $\alpha$ yields
\begin{center}
	\begin{tikzpicture}[scale=0.8] 
	\begin{scope}[xshift=0cm,yshift=1cm]
	\node at (0, 0)   (d) {$\displaystyle\sum_\alpha$};
	\end{scope}
	\begin{scope}[xshift=.5cm,yshift=-.5cm]
	\draw (0,3)--(3.8,3);
	\draw (0,2)--(3.8,2);
	\draw (0,1)--(3.8,1);
	\draw (0,0)--(3.8,0);
	\node[draw,ellipse,scale=.9,fill=white,minimum height=3cm] at (.8, 1.5)   (c) {$S_4$};
	\node[draw,circle,minimum height=.9cm,scale=.8,fill=white] at (2, 3)   (b) {$\lambda^\alpha$};
	\node[draw,circle,minimum height=.9cm,scale=.8,fill=white] at (3, 3)   (b) {$\lambda^\alpha$};
	\end{scope}
	\begin{scope}[xshift=4.9cm,yshift=1cm]
	\node at (0, 0)   (d) {$+$};
	\end{scope}
	\begin{scope}[xshift=5.5cm,yshift=-.5cm]
	\draw (0,3)--(2.8,3);
	\draw (0,2)--(2.8,2);
	\draw (0,1)--(2.8,1);
	\draw (0,0)--(2.8,0);
	\node[draw,ellipse,scale=.9,fill=white,minimum height=3cm] at (.8, 1.5)   (c) {$S_4$};
	\node[draw,circle,minimum height=.9cm,scale=.8,fill=white] at (2, 3)   (b) {$\lambda^\alpha$};
	\node[draw,circle,minimum height=.9cm,scale=.8,fill=white] at (2, 2)   (b) {$\lambda^\alpha$};
	\end{scope}
	\begin{scope}[xshift=10cm]
	\begin{scope}[xshift=0cm,yshift=1cm]
	\node at (-.75, 0)   (d) {$\displaystyle=\sum_\alpha$};
	\end{scope}
	\begin{scope}[xshift=0cm,yshift=-.5cm]
	\draw (.2,3)--(3.8,3);
	\draw (.2,2)--(3.8,2);
	\draw (.2,1)--(3.8,1);
	\draw (.2,0)--(3.8,0);
	\node[draw,ellipse,scale=.9,fill=white,minimum height=3cm] at (2, 1.5)   (c) {$S_4$};
	\node[draw,circle,minimum height=.9cm,scale=.8,fill=white] at (.8, 1)   (b) {$\lambda^\alpha$};
	\node[draw,circle,minimum height=.9cm,scale=.8,fill=white] at (3, 3)   (b) {$\lambda^\alpha$};
	\end{scope}
	\begin{scope}[xshift=4.5cm,yshift=1cm]
	\node at (0, 0)   (d) {$+$};
	\end{scope}
	\begin{scope}[xshift=5cm,yshift=-.5cm]
	\draw (.2,3)--(3.8,3);
	\draw (.2,2)--(3.8,2);
	\draw (.2,1)--(3.8,1);
	\draw (.2,0)--(3.8,0);
	\node[draw,ellipse,scale=.9,fill=white,minimum height=3cm] at (2, 1.5)   (c) {$S_4$};
	\node[draw,circle,minimum height=.9cm,scale=.8,fill=white] at (1, 0)   (b) {$\lambda^\alpha$};
	\node[draw,circle,minimum height=.9cm,scale=.8,fill=white] at (3, 3)   (b) {$\lambda^\alpha$};
	\end{scope}
	\end{scope}
	\end{tikzpicture}
\end{center}
Exploiting the previously derived identities allows us to replace the summations with rewirings:
\begin{center}
	\begin{tikzpicture}[scale=0.8] 
	\begin{scope}[xshift=0cm,yshift=1cm]
	\node at (0, 0)   (d) {$d$};
	\end{scope}
	\begin{scope}[xshift=.5cm,yshift=-.5cm]
	\draw (0,3)--(4,3);
	\draw (0,2)--(4,2);
	\draw (0,1)--(4,1);
	\draw (0,0)--(4,0);
	\node[draw,ellipse,scale=.9,fill=white,minimum height=3cm] at (.8, 1.5)   (c) {$S_4$};
	\end{scope}
	\begin{scope}[xshift=4.95cm,yshift=1cm]
	\node at (0, 0)   (d) {$+$};
	\end{scope}
	\begin{scope}[xshift=5.5cm,yshift=-.5cm]
	\draw (0,3)--(2,3);
	\draw (0,2)--(2,2);
	\draw (0,1)--(4,1);
	\draw (0,0)--(4,0);
	\node[draw,ellipse,scale=.9,fill=white,minimum height=3cm] at (.8, 1.5)   (c) {$S_4$};
	\begin{scope}[xshift=2cm,yshift=2cm]
	\draw[dark] (0,0) to[out=0,in=240](1,.5)to[out=70,in=180](2,1);
	\draw[line width=3pt, white] (0,1)to[out=0,in=90](1,.5)to[out=270,in=180](2,0);
	\draw[dark] (0,1)to[out=0,in=110](1,.5)to[out=290,in=180](2,0);
	\end{scope}
	\end{scope}
	\begin{scope}[xshift=11cm]
	\begin{scope}[xshift=0cm,yshift=1cm]
	\node at (-.75, 0)   (d) {$=$};
	\end{scope}
	\begin{scope}[xshift=0cm,yshift=-.5cm]
	\draw (0,3)--(3,3);
	\draw (3.75,3)--(4,3);
	\draw (0,2)--(4,2);
	\draw (0,1)--(.25,1);
	\draw (1,1)--(4,1);
	\draw (0,0)--(4,0);
	\node[draw,ellipse,scale=.9,fill=white,minimum height=3cm] at (2, 1.5)   (c) {$S_4$};
	\draw[line width=3pt, white] (3,3)to[out=0,in=90](3.3,2.8)to[out=270,in=30](3,2.5)--(1,1.5)to[out=210,in=90](0.7,1.2)to[out=270,in=180](1,1);
	\draw[dark] (3,3)to[out=0,in=90](3.3,2.8)to[out=270,in=30](3,2.5)--(1,1.5)to[out=210,in=90](0.7,1.2)to[out=270,in=180](1,1);
	\draw[line width=3pt, white] (.25,1)to[out=0,in=180](1.5,.5)to[out=0,in=270](3.5,2)to[out=90,in=180](3.75,3);
	\draw[dark] (.25,1)to[out=0,in=180](1.5,.5)to[out=0,in=270](3.5,2)to[out=90,in=180](3.75,3);
	\node at (2, -1)   (d) {$(i)$};
	\end{scope}
	\begin{scope}[xshift=4.5cm,yshift=1cm]
	\node at (0, 0)   (d) {$+$};
	\end{scope}
	\begin{scope}[xshift=5cm,yshift=-.5cm]
	\draw (0,3)--(3,3);
	\draw (3.75,3)--(4,3);
	\draw (0,2)--(4,2);
	\draw (0,1)--(4,1);
	\draw (0,0)--(.25,0);
	\draw (1,0)--(4,0);
	\node[draw,ellipse,scale=.9,fill=white,minimum height=3cm,text depth=1.2cm] at (2, 1.5)   (c) {$S_4$};
	\draw[line width=3pt, white] (.25,0)to[out=0,in=180](1.5,.5)to[out=0,in=270](3.5,2)to[out=90,in=180](3.75,3);
	\draw[dark] (.25,0)to[out=0,in=180](1.5,.5)to[out=0,in=270](3.5,2)to[out=90,in=180](3.75,3);
	\draw[line width=3pt, white] (3,3)to[out=0,in=90](3.2,2.8)to[out=270,in=50](3,2.5)--(1,0.5)to[out=230,in=90](0.8,0.2)to[out=270,in=180](1,0);
	\draw[dark] (3,3)to[out=0,in=90](3.2,2.8)to[out=270,in=50](3,2.5)--(1,0.5)to[out=230,in=90](0.8,0.2)to[out=270,in=180](1,0);
	\node at (2, -1)   (d) {$(ii)$};
	\end{scope}
	\end{scope}
	\end{tikzpicture}
\end{center}
The LHS can be factorised slightly to obtain
\begin{center}
	\begin{tikzpicture}[scale=0.8] 
	\begin{scope}[xshift=.5cm,yshift=-.5cm]
	\draw (0,3)--(4,3);
	\draw (0,2)--(4,2);
	\draw (0,1)--(4,1);
	\draw (0,0)--(4,0);
	\node[draw,ellipse,scale=.9,fill=white,minimum height=3cm] at (.8, 1.5)   (c) {$S_4$};
	\node[draw,circle,minimum height=1.2cm,scale=.9,fill=white] at (2.5, 2.5)   (b) {$M$};
	\end{scope}
	\begin{scope}[xshift=5.40cm,yshift=1cm]
	\node at (0, 0)   (d) {$\displaystyle=$};
	\end{scope}
	\begin{scope}[xshift=6.25cm,yshift=-.5cm]
	\draw (0,3)--(4,3);
	\draw (0,2)--(4,2);
	\draw (0,1)--(4,1);
	\draw (0,0)--(4,0);
	\node[draw,ellipse,scale=.9,fill=white,minimum height=3cm] at (.8, 1.5)   (c) {$S_4$};
	\end{scope}
	\begin{scope}[xshift=10.75cm,yshift=1cm]
	\node at (0, 0)   (d) {$\displaystyle+\frac{1}{d}$};
	\end{scope}
	\begin{scope}[xshift=11.5cm,yshift=-.5cm]
	\draw (0,3)--(2,3);
	\draw (0,2)--(2,2);
	\draw (0,1)--(4,1);
	\draw (0,0)--(4,0);
	\node[draw,ellipse,scale=.9,fill=white,minimum height=3cm] at (.8, 1.5)   (c) {$S_4$};
	\begin{scope}[xshift=2cm,yshift=2cm]
	\draw[dark] (0,0) to[out=0,in=240](1,.5)to[out=70,in=180](2,1);
	\draw[line width=3pt, white] (0,1)to[out=0,in=90](1,.5)to[out=270,in=180](2,0);
	\draw[dark] (0,1)to[out=0,in=110](1,.5)to[out=290,in=180](2,0);
	\end{scope}
	\end{scope}
	\end{tikzpicture}
\end{center}
where
\begin{center}
	\begin{tikzpicture}[scale=0.8] 
	\begin{scope}[xshift=-2cm]
	\draw (0,1)--(2,1);
	\draw (0,0)--(2,0);
	\node[draw,ellipse,scale=.9,fill=white,minimum height=1.5cm] at (1, .5)   (c) {$M$};
	\end{scope}
	\begin{scope}[xshift=.5cm,yshift=.5cm]
	\node[] at (0,0)   (a) {$=$};
	\end{scope}
	\begin{scope}[xshift=1cm]
	\draw (0,1)--(1,1);
	\draw (0,0)--(1,0);
	\end{scope}
	\begin{scope}[xshift=2.75cm,yshift=.5cm]
	\node[] at (0,0)   (a) {$\displaystyle+\frac{1}{d}$};
	\end{scope}
	\begin{scope}[xshift=3.5cm]
	\draw (0,0) to[out=0,in=240](1,.5)to[out=70,in=180](2,1);
	\draw[line width=3pt, white] (0,1)to[out=0,in=90](1,.5)to[out=270,in=180](2,0);
	\draw (0,1)to[out=0,in=110](1,.5)to[out=290,in=180](2,0);
	\end{scope}
	\end{tikzpicture}
\end{center}
The inverse of $M$ is given by
\begin{center}
	\begin{tikzpicture}[scale=0.8] 
	\begin{scope}[xshift=-2.2cm]
	\draw (0,1)--(2,1);
	\draw (0,0)--(2,0);
	\node[draw,ellipse,scale=.9,fill=white,minimum height=1.5cm] at (1, .5)   (c) {$M^{-1}$};
	\end{scope}
	\begin{scope}[xshift=1cm,yshift=.5cm]
	\node[] at (0,0)   (a) {$\displaystyle=\frac{d^2}{d^2-1}$};
	\end{scope}
	\begin{scope}[xshift=2.25cm]
	\draw (0,1)--(1,1);
	\draw (0,0)--(1,0);
	\end{scope}
	\begin{scope}[xshift=4.5cm,yshift=.5cm]
	\node[] at (0,0)   (a) {$\displaystyle-\frac{d}{d^2-1}$};
	\end{scope}
	\begin{scope}[xshift=5.5cm]
	\draw (0,0) to[out=0,in=240](1,.5)to[out=70,in=180](2,1);
	\draw[line width=3pt, white] (0,1)to[out=0,in=90](1,.5)to[out=270,in=180](2,0);
	\draw (0,1)to[out=0,in=110](1,.5)to[out=290,in=180](2,0);
	\end{scope}
	\end{tikzpicture}
\end{center}
Multiplying both sides on the right by $M^{-1}\otimes \mathds{1}\otimes \mathds{1}$ gives
\begin{center}
	\begin{tikzpicture}[scale=0.8] 
	\begin{scope}[xshift=.5cm,yshift=-.5cm]
	\draw (0,3)--(2,3);
	\draw (0,2)--(2,2);
	\draw (0,1)--(2,1);
	\draw (0,0)--(2,0);
	\node[draw,ellipse,scale=.9,fill=white,minimum height=3cm] at (1, 1.5)   (c) {$S_4$};
	\end{scope}
	\begin{scope}[xshift=5cm,yshift=-.5cm]
	\node at (-1.25, 1.5)   (d) {$\displaystyle=\frac{1}{d^2-1}$};
	\draw[line width=3pt, white] (0,1)--(2,3);
	\draw (0,1)--(2,3);
	\draw[line width=3pt, white] (0,0)--(2,2);
	\draw (0,0)--(2,2);
	\draw[line width=3pt, white] (0,3)--(2,1);
	\draw (0,3)--(2,1);
	\draw[line width=3pt, white] (0,2)--(2,0);
	\draw (0,2)--(2,0);
	\end{scope}
	\begin{scope}[xshift=9.5cm,yshift=-.5cm]
	\node at (-1.3, 1.5)   (d) {$\displaystyle-\frac{1}{d(d^2-1)}$};
	\draw[line width=3pt, white] (0,2)--(2,0);
	\draw (0,2)--(2,0);
	\draw[line width=3pt, white] (0,1)--(2,2);
	\draw (0,1)--(2,2);
	\draw[line width=3pt, white] (0,0)--(2,3);
	\draw (0,0)--(2,3);
	\draw[line width=3pt, white] (0,3)--(2,1);
	\draw (0,3)--(2,1);
	\end{scope}
	\begin{scope}[xshift=13.75cm,yshift=-.5cm]
	\node at (-1.25, 1.5)   (d) {$\displaystyle+\frac{1}{d^2-1}$};
	\draw[line width=3pt, white] (0,1)to[bend right=40](2,2);
	\draw (0,1)to[bend right=40](2,2);
	\draw[line width=3pt, white] (0,2)to[bend left=40](2,1);
	\draw (0,2)to[bend left=40](2,1);
	\draw[line width=3pt, white] (0,0)--(2,3);
	\draw (0,0)--(2,3);
	\draw[line width=3pt, white] (0,3)--(2,0);
	\draw (0,3)--(2,0);
	\end{scope}
	\begin{scope}[xshift=18.5cm,yshift=-.5cm]
	\node at (-1.5, 1.5)   (d) {$\displaystyle-\frac{1}{d(d^2-1)}$};
	\draw[line width=3pt, white] (0,1)--(2,3);
	\draw (0,1)--(2,3);
	\draw[line width=3pt, white] (0,0)--(2,2);
	\draw (0,0)--(2,2);
	\draw[line width=3pt, white] (0,2)to[bend right=40](2,1);
	\draw (0,2)to[bend right=40](2,1);
	\draw[line width=3pt, white] (0,3)--(2,0);
	\draw (0,3)--(2,0);
	\end{scope}
	\end{tikzpicture}
\end{center}
In deriving this equation we used the following identity for (i):
\begin{center}
	\begin{tikzpicture}[scale=0.8] 
	\node at (-.5,0)   (d) {$\displaystyle (i)=\int du$};
	\begin{scope}[xshift=1cm,yshift=-1.5cm]
	\draw (0,3)--(2,3);
	\draw (0,2)--(4,2);
	\draw (1,1)--(4,1);
	\draw (0,0)--(4,0);
	\draw[line width=3pt, white] (0,1)to[out=0,in=180](1,.5)to[out=0,in=180](4,3);
	\draw[dark] (0,1)to[out=0,in=180](1,.5)to[out=0,in=180](4,3);
	\draw[line width=3pt, white] (2,3)to[out=0,in=0](1,1.5)to[out=180,in=180](.4,1)--(1,1);
	\draw[dark] (2,3)to[out=0,in=0](1,1.5)to[out=180,in=180](.4,1)--(1,1);
	\node[draw,circle,minimum height=.9cm,scale=.7,fill=white] at (1, 3)   (b) {$U^\dagger$};
	\node[draw,circle,minimum height=.9cm,scale=.7,fill=white] at (1, 2)   (b) {$U^\dagger$};
	\node[draw,circle,minimum height=.9cm,scale=.7,fill=white] at (1, 1)   (b) {$U$};
	\node[draw,circle,minimum height=.9cm,scale=.7,fill=white] at (1, 0)   (b) {$U$};
	\end{scope}
	\node at (5.5,0)   (d) {$=$};
	\begin{scope}[xshift=6cm,yshift=-1.5cm]
	\draw[dark] (0,2)--(4,2);
	\draw[dark] (0,0)--(4,0);
	\node[draw=dark,ellipse,scale=.9,fill=white,minimum height=2cm,text depth=1.2cm] at (3, 1)   (c) {$S_2$};
	\draw[line width=3pt, white](0,3)to[out=0,in=110](1,2)to[out=290,in=180](2,1)--(4,1);
	\draw (0,3)to[out=0,in=110](1,2)to[out=290,in=180](2,1)--(4,1);
	\draw[line width=3pt, white] (0,1) to[out=0,in=240](1,2)to[out=70,in=180](2,3)--(4,3);
	\draw (0,1) to[out=0,in=240](1,2)to[out=70,in=180](2,3)--(4,3);
	\end{scope}
	\node at (10.5,0)   (d) {$=$};
	\begin{scope}[xshift=11cm,yshift=-1.5cm]
	\draw[line width=3pt, white] (0,1)--(2,3);
	\draw (0,1)--(2,3);
	\draw[line width=3pt, white] (0,0)--(2,2);
	\draw (0,0)--(2,2);
	\draw[line width=3pt, white] (0,3)--(2,1);
	\draw (0,3)--(2,1);
	\draw[line width=3pt, white] (0,2)--(2,0);
	\draw (0,2)--(2,0);
	\end{scope}
	\end{tikzpicture}
\end{center}
(A similar argument was also exploited for (ii).)

The explicit representation for $S_4$ allows us to derive the final line for the expression of the quantum risk Eq.~(\ref{eq:quantumrisk}) according to
\begin{center}
	\begin{tikzpicture}[scale=0.8] 
		\node at (.15,0)   (d) {$\displaystyle\int du \bra{0}U^\dagger X^\dag U \ket{0}\bra{0}U^\dagger X U \ket{0}\displaystyle=\int du$};
	\begin{scope}[xshift=5.5cm,yshift=-1.5cm]
	\draw (0,3)--(2,3);
	\draw (0,2)--(2,2);
	\draw (1,1)--(2,1);
	\draw (1,0)--(2,0);
	\draw[dark] (2,2)--(3,2);
	\draw[dark] (3,1.5) arc(-90:90:.25);
	\draw[dark] (0,1.5)--(3,1.5);
	\draw[dark] (0,1.5) arc(90:270:.75);
	\draw[dark] (0,0)--(1,0);
	\draw[dark] (2,3)--(3,3);
	\draw[dark] (3,2.5) arc(-90:90:.25);
	\draw[dark] (0,2.5)--(3,2.5);
	\draw[white, line width=3pt] (0,2.5) arc(90:270:.75);
	\draw[dark] (0,2.5) arc(90:270:.75);
	\draw[dark] (0,1)--(1,1);

	\node[draw,circle,minimum height=.9cm,scale=.7,fill=white] at (1, 3)   (b) {$U^\dagger$};
	\node[draw,circle,minimum height=.9cm,scale=.7,fill=white] at (1, 2)   (b) {$U^\dagger$};
	\node[draw,circle,minimum height=.9cm,scale=.7,fill=white] at (1, 1)   (b) {$U$};
	\node[draw,circle,minimum height=.9cm,scale=.7,fill=white] at (1, 0)   (b) {$U$};
	\node[draw=dark,circle,minimum height=.9cm,scale=.7,fill=white] at (0, 3)   (b) {$\bra{0}$};
	\node[draw=dark,circle,minimum height=.9cm,scale=.7,fill=white] at (0, 2)   (b) {$\bra{0}$};
	\node[draw=dark,circle,minimum height=.9cm,scale=.7,fill=white] at (2, 3)   (b) {$X^\dag$};
	\node[draw=dark,circle,minimum height=.9cm,scale=.7,fill=white] at (2, 2)   (b) {$X$};
	\node[draw=dark,circle,minimum height=.9cm,scale=.7,fill=white] at (2, 1)   (b) {$\ket{0}$};
	\node[draw=dark,circle,minimum height=.9cm,scale=.7,fill=white] at (2, 0)   (b) {$\ket{0}$};
	\end{scope}
	\end{tikzpicture}
\end{center}

\subsection{Classical NFL theorem for invertible functions}

A classical NFL-like theorem for invertible functions can be readily obtained by adapting the original argument. We first assume the cardinalities $|X|$ and $|Y|$ are equal, so that invertibility can be translated to surjectivity and injectivity. One can take both these properties into account by requiring that for each element $x \in X \setminus S$ the image under the hypothesis is in the complement of the image of the training set. Defining the image of $S$ under $h_S$ as $h_S(S) := \{y \in Y | y = h_S(z) , z \in S \}$, we have that $h_S(x) \notin  h_S(S)$, $\forall x \in X \setminus S$. Thus when optimising with $n$ training points, the cardinality of the possible image of $h_S$ when confronted with a point not in the training set is now $|Y| - n = |X| - n$. This is equivalent to optimising with the output set $\left(h_S(S)\right)^{C}$, so the complement of $h_S(S)$ in $Y$. Inserting this into the original statement \ref{eq:original_NFL_statement} then yields a theorem for invertible functions and hypotheses $f$ and $h_S$:
\begin{equation}
	\eqbox{
	\mathbb{E}_f\left[\mathbb{E}_S\left[R_f(h_S)\right]\right]\ge \left(1-\frac{1}{|X| - n}\right)\left(1-\frac{n}{|X|}\right) = \frac{|X| - (n+1)}{|X| - n}\frac{|X| - n}{|X|} = 1 - \frac{n+1}{|X|}.}
\end{equation}

\end{document}